\documentclass[superscriptaddress,twocolumn,secnumarabic,amssymb, nobibnotes, aps, prd]{revtex4-1}

\usepackage{graphicx,array,multirow}
\begin{document}

\title{Neutrino energy reconstruction from one muon and one proton events}%

\author{Andrew P. Furmanski}
\affiliation{
The University of Manchester, Oxford Road, Manchester, M13 9PL, UK}
\affiliation{
Fermi National Accelerator Laboratory, Batavia, Illinois 60510, USA}
\author{Jan T. Sobczyk}%
\affiliation{
Fermi National Accelerator Laboratory, Batavia, Illinois 60510, USA}
\affiliation{Institute of Theoretical Physics, University of Wroc\l aw, pl. M. Borna 9,
50-204, Wroc\l aw, Poland}
\date{\today}%
\begin{abstract}
We propose a new method of selection of high purity charge current quasielastic neutrino events with a good reconstruction of interacting neutrino energy. Performance of the method was verified with several tests using GENIE, NEUT and NuWro Monte Carlo events generators with carbon and argon targets. The method can be useful in neutrino oscillation studies with a few GeV energy beams.
\end{abstract}

\maketitle

\section{Introduction}
\label{sec:intro}

A pattern of neutrino oscillations is neutrino energy dependent. It is why in recent years there has been a lot of discussion on unbiased neutrino energy
reconstruction based on particles detected in the final state after escaping target nucleus. The topic is important because neutrino flux is never
mono-energetic and even in off-axis configuration has a significant spread. The discussion has became even more intense after realizing that in the few GeV 
energy region (where most current and planned long baseline oscillation experiments work) there is a significant contribution from two-body current mechanism
making the situation even more complicated \cite{AguilarArevalo:2010zc, Martini:2009uj, *Nieves:2011pp, Martini:2012fa, *Lalakulich:2012hs, *Nieves:2012yz}, for review articles see Ref. \cite{Morfin:2012kn, *Alvarez-Ruso:2014bla}.

In the few GeV region the most important reaction is charge current quasi-elastic (CCQE) scattering:

\begin{eqnarray}
\label{eq:ccqe}
\nu_\mu \ n \rightarrow \mu^-\ p\\
\bar\nu_\mu \ p \rightarrow \mu^+\ n.
\end{eqnarray}
In what follows we focus on muon neutrinos only.
CCQE is well defined only in case of free nucleon scattering. In oscillation experiments targets are mostly nuclei. If momentum and energy
transfered to the hadronic system is large enough, one can rely on impulse approximation (IA) \cite{Benhar:2005dj} according to which
interactions occur on individual (bound and moving)
nucleons and the meaning of CCQE remains valid. This is confirmed by a clear observation of QE peak in electron nucleus target inclusive cross section measurements.
In neutrino experiments analogous identification of the QE peak is impossible, because neutrino beams are never monoenergetic so one cannot know the 
energy and momentum transfer on an event by event basis (for an attempt to estimate energy and momentum transfer in neutrino experiment see Ref.  \cite{Rodrigues:2015hik}).

In the IA picture neutrino scattering is a two step process and a primary interaction on a bound nucleon 
(or nucleon pair) is followed by hadron rescatterings called final state interactions (FSI). 
This is a reason why neutrino experimental groups introduced a notion of CCQE-like events. They are
defined as those with a muon and no mesons in the final state. The advantage of this notion is that it is defined with no ambiguity and also it is relatively close to genuine
CCQE in the IA picture. A widely discussed MiniBooNE measurement of CCQE double differential cross section \cite{AguilarArevalo:2010zc} was in fact a CCQE-like measurement after the results from tables VI and VIII in Ref. \cite{AguilarArevalo:2010zc} are added.  A more recent measurement of CCQE-like cross section was reported by T2K \cite{Abe:2016tmq}. However, CCQE-like samples of events contain also those in which a real pion is first produced and then absorbed, and also those arising from two-body
mechanism, typically leading to two nucleon knock-out.
Notice that FSI effects 
can also make a CCQE event to be not CCQE-like because there is
a small probability that nucleon produces a pion before leaving a nucleus. It should become clear that precise description of neutrino interactions
in the few GeV energy region requires a careful combination of several dynamical models. This is a motivation for numerous neutrino Monte Carlo (MC) event generators studies \cite{PDG-MC}. 

The goal of this paper is to propose a novel method to select high purity CCQE events out of CCQE-like experimental samples. The method works for events with exactly two particles detected in the final state: muon and proton.
Muon identification is usually quite simple, however reconstructing the short tracks left by protons is more challenging.
Nevertheless, in recent years experimental groups like T2K, MINERvA, ArgoNeUT presented various studies using information about the reconstructed proton, see e.g. Ref. \cite{Walton:2014esl}. 
In the case of a proton the ability to measure it depends on details of the detector technique. In segmented scintillators used by T2K and MINERvA a typical threshold for
proton identification is $\sim 400$~MeV/c. However, in liquid argon used in ArgoNeuT, MicroBooNE and proposed in the DUNE experiment 
the threshold is much lower: $\sim 200$~MeV/c. 

A sample with no mesons, exactly one muon and one proton consists of four basic categories of events: (i) CCQE events with a proton that did not suffer from FSI effects; (ii) CCQE events with a proton that at least once rescattered before was knocked-out and detected; (iii) $\pi$ production events with subsequent $\pi$ absorption; (iv) two-body current events. The relative sizes of contributions from these four categories depend on neutrino beam spectrum, target nucleus and threshold for proton track identification. We are going to propose a selection criterium which eliminates a vast majority of the events from categories (iii) and (iv) keeping most of events from categories (i) and (ii). The relative population of subsamples (i) and (ii) depends on the probability that a proton goes through nucleus without interacting. This probability is often expressed as nuclear transparency. Clearly, transparency depends on nucleus size and is larger for carbon (about $65\%$) than it is for argon. We are not aware of any proton transparency measurement for argon. More details on transparency studies for targets like iron can be found in Ref. \cite{Garino:1992ca, *O'Neill:1994mg, *Hen:2012yva}.

Our argument is rather elementary and is based merely on energy and momentum conservation. If the event was indeed CCQE with no proton final state interactions, energy and momentum conservation allow to resolve the kinematics of the process completely and calculate the initial neutrino and neutron three-momenta. Recently, various transverse kinematics studies were done exploring momentum conservation in the plane perpendicular to the neutrino
momentum vector \cite{Lu:2015tcr, *Lu:2016mjf}. In our computations we use also information from momentum conservation along the neutrino momentum vector. The obtained values of neutron momentum and neutrino energy should be almost exact for events from the category (i). For the events from the category (ii) exactness of the obtained values depends on how severe were rescatterings experienced by the proton. For the events from the categories (iii) and (iv) our procedure is expected to produce to much extent random numbers. It is because the very assumption that the interaction occured on a neutron is not satisfied and the interaction mechanism is more complicated. This is confirmed by MC simulations. 
Appropriate cuts will be able to eliminate most of them. 

As said before, the goal is to select true CCQE events. Target neutron momenta and separation energies can be described by means of spectral function (SF) \cite{Benhar:1994hw}. Neutron momentum distribution consists of two parts: mean field part dominating for $p<250$~MeV/c and correlation part responsible for the high momentum tail $p>250$~MeV/c \cite{Hen:2014nza}. The mean field part can be understand in a language of the shell model. The correlation part comes from short range correlated pairs. In the few GeV neutrino energy region a vast majority of CCQE events comes from interactions on neutrons described by the mean field. Our selection aims to select those events by requiring a reconstructed neutron momentum of the order of $\sim 250$~MeV or lower. In our computations we will have to estimate, on event by event basis, excitatation energy of the remnant nuclues after a proton was knocked out. The estimation will be consistent with the assumption that we select mean field neutrons. Relevant information will be given in terms of removal energy from individual neutron energy levels. 

Our main result is that the performance of the proposed selection method is indeed very good. We did numerical computations using three different MC event generators: GENIE \cite{Andreopoulos:2009rq}, NEUT \cite{Hayato:2009zz} and NuWro \cite{Golan:2012wx}. Each of them treats nuclear effects and in particular FSI effects in different way, yet the conclusions we obtained are always quite similar. Thus we infer that a high purity CCQE sample of events can be obtained. For the selected sample we show that the
precision of the energy reconstruction is extremely good, and outperforms other available methods.

Our paper is organized as follows. In Sect. \ref{sec:one} we present a framework in which the neutron momentum and neutrino energy are evaluated.
In Sect. \ref{sec:results} we demonstrate the performance of the proposed method. We will discuss several examples using MiniBooNE, T2K, and NuMI $\nu_\mu$ fluxes with interactions occuring on carbon or argon. Sects. \ref{sec:discussion} and \ref{sec:conclusions} contain discussion and final conclusions.

\section{Model}
\label{sec:one}

Suppose that a CCQE interaction occured on a nucleon inside nucleus, one nucleon is knocked out and detected together with a final state muon. Suppose also that 
target nucleus is at rest and no other particles are knocked out.

The energy and momentum conservation read:

\begin{eqnarray}
\label{eq:conservation}
E + M_A =& E' + E_{p'} + E_{A-1}\\
\vec k =& \vec k' + \vec p' + \vec p_{A-1}
\end{eqnarray}
where $(E,\vec k)$, $(E',\vec k')$ are the neutrino and muon four-momenta, $M_A$ is the target nucleus mass, $(E_{A-1}, \vec p_{A-1})$, $(E_{p'}, \vec p')$ are the final state nucleus and final state nucleon four-momenta.

In the IA picture, the interaction occurs on a nucleon with momentum $\vec p$. If no final state interactions took place it must be that

\begin{eqnarray}
\label{eq:zeromomentum}
\vec p = - \vec p_{A-1}.
\end{eqnarray}
this means that the initial nucleus state is factorized into a nucleon participating in the interaction and a spectator remnant nucleus with $A-1$ nucleons.

We can decompose $\vec p$ into components parallel and perpendicular with respect to the neutrino direction, $\vec k$, which is assumed to be known:

\begin{eqnarray}
\label{eq:decomposition}
\vec p = \vec p_L + \vec p_T
\end{eqnarray}

The same can be done with vectors $\vec k'$ and $\vec p'$. We get the following equations:

\begin{eqnarray}
\label{eq:conservation2}
E + M_A =& E' + E_{p'} + E_{A-1}\\
E =& k'_L + p'_L - p_L\\
0 =& \vec k'_T + \vec p'_T -\vec p_T.
\end{eqnarray}
We used the fact that for neutrino $E=|\vec k|$. $p_L$, $k'_L$ and $p'_L$ denote projections of the corresponding three vectors on the direction of $\vec k$.

The final state nucleus is in general in an excited state and its invariant mass is $M_{A-1}^*$. 
If the final state muon and proton are measured, $\vec p_T$ is known and we obtain two equations for $E$ and $p_L$ that can be easily solved. We get:

\begin{eqnarray}
\label{eq:solution_p}
p_L=\frac{ (M_A + k'_L + p'_L - E' - E_{p'})^2 - p_T^2 - {M_{A-1}^*}^2}{ 2(M_A + k'_L + p'_L - E' - E_{p'}) }
\end{eqnarray}

\begin{eqnarray}
\label{eq:full_p}
p = \sqrt{\vec p_T^2 + p_L^2},
\end{eqnarray}

\begin{eqnarray}
\label{eq:solution_E}
E= k'_L + p'_L - p_L.
\end{eqnarray}

We notice that an anologous derivation can be done in the IA scheme
treating a hit nucleon as a bound one with an unknown binding energy, forgetting about initial and final state nuclei. Two derivations can be made completely
equivalent if the binding energy is treated as nucleon momentum dependent.

We will not assume a constant average value of $M_{A-1}^*$. Instead we introduce a probability distribution for the final state excitation energy using information about neutron occupancy in given nucleus. In this paper we will discuss $^{12}$C and $^{40}$Ar. In the case of argon relevant information can be found in \cite{Ankowski:2007uy}. For the neutron seperation energy $E$ Gaussian distribution is assumed with central values $E_\alpha$ and standard deviation $\sigma_\alpha$, see Table \ref{table:argon}. 

For carbon, we used information about proton separation energies and widths from \cite{Frullani:1984nn}. Following \cite{Ankowski:2012ei} we accounted for a fact that in carbon neutrons are most deeply bound by $\Delta \approx 2.76$~MeV. The information is collected in Table \ref{tab:carbon}.

To summarize, we select separation energy using the probability distribution 

\begin{eqnarray}
\label{eq:prob-distr}
P(E)=\frac{1}{N}\sum_\alpha n_\alpha G(E-E_\alpha, \sigma_\alpha)
\end{eqnarray}
where $N$ is a number of neutrons 

\[ \sum_\alpha n_\alpha = N\]
and $G(x-x_0,\sigma)$ is a Gaussian distribution with mean value $x_0$ and standard deviation $\sigma$. The binding energies of carbon and argon nuclei in the ground state are respectively: $B=92.16$~MeV and $B=343.81$~MeV.
It means that in the numerical computations for carbon we take

\[ M_A= 6M_n+6M_p - 92.16~{\rm MeV},\]
\[ M^*_{A-1}= M_A-M_n+E\]
where $M_p$ and $M_n$ denote proton and neutron mass. Anologous formulas are used for argon.

\begin{table}[h!]
  \centering
  \begin{tabular}{l|c|c|c}
    Subshell & $E_\alpha$ [MeV] & $\sigma_\alpha$ [MeV] & \# neutrons $n_\alpha$\\
    \hline
    $1s_{1/2}$ & 62 & 6.25 & 2\\
    $1p_{3/2}$ & 40 & 3.75 & 4\\
    $1p_{1/2}$ & 35 & 3.75 & 2\\
    $1d_{5/2}$ & 18 & 1.25 & 6\\
    $2s_{1/2}$ & 13.15 & 1 & 2\\
    $1d_{3/2}$ & 11.45 & 0.75 & 4\\
    $1f_{7/2}$ & 5.56 & 0.75 & 2
  \end{tabular}
  \caption{Neutron shell structure in $^{40}$Ar \cite{Ankowski:2007uy}}
   \label{table:argon}
\end{table}

\begin{table}[h!]
  \centering
  \begin{tabular}{l|c|c|c}
    Subshell & $E_\alpha$ [MeV] & $\sigma_\alpha$ [MeV] & \# neutrons $n_\alpha$\\
    \hline
    $1s_{1/2}$ & 40.8 & 9.1 & 2\\
    $1p_{3/2}$ & 20.3 & 5 & 4
  \end{tabular}
  \caption{Neutron shell structure in $^{12}$C \cite{Frullani:1984nn}}
   \label{tab:carbon}
\end{table}

\subsection{Monte Carlo generators}

In our tests we used three widely used MC generators: GENIE \cite{Andreopoulos:2009rq}, NEUT \cite{Hayato:2009zz} and NuWro \cite{Golan:2012wx}. Their basic structure is similar, however they differ in many details that play a role in the comparisons we did. 

In CCQE events target neutrons are typically described with (local) Fermi gas model. GENIE uses Bodek-Ritchie modification of the Fermi gas with a large momentum tail added in the neutron momentum distribution accounting for nucleon-nucleon correlation effects \cite{Bodek:1980ar}. One of the options in NuWro is to use spectral function \cite{Benhar:1989aw}. In this approach one distinguishes if an interaction occurs on a nucleon discribed by a shell model or on a nucleon forming a correlated pair. In the second case the existence of a correlated spectator nucleon is assumed,  that also propagates through nucleus. Its initial momentum is postulated to be opposite (as a three-vector) to that of the interacting nucleon. 

For RES events (mostly single pion production) NEUT and GENIE use Rein-Sehgal model \cite{Rein:1980wg} with upgraded information about resonance properties. NEUT includes resonance interference effects and anisotropy of the distribution of pions resulting from $\Delta$ decays \cite{Radecky:1981fn, *Barish:1978pj}. NuWro has a separate treatment of the $\Delta$ resonance with form factors fitted to the experimental data. Heavier resonances are included only in an approximate way using quark-hadron duality arguments. All three MCs differ in a way in which non-resonant background contribution is included. All three  MCs account for $\Delta$ in-medium self-energy \cite{Oset:1987re} using different approximations.
NuWro models $\Delta$ finite life-time inside nucleus. 

More inelastic events (DIS) treatments are more similar \cite{Bodek:2002vp}. The same model of inclusive DIS crosss section is used and then PYTHIA fragmentation routines produce the final states. Differences are in the values of some PYTHIA parameters and also in a kinematical region where this formalism is used. NuWro extends it to $W=1.6$~GeV while GENIE and NEUT only to $1.8$ or $2.0$ GeV with KNO scaling arguments \cite{Koba:1972ng} applied in the transition region.

In the versions of NEUT (v. 5.1.4.2) and GENIE GENIE (v. 2.8.6) used in this study two-body current ("MEC") events are not produced. NuWro uses as a default Nieves model \cite{Nieves:2011pp} with a momentum transfer cut $q\leq 1.2$~GeV/c \cite{Gran:2013kda}. $85\%$ of MEC events occur on proton-neutron pairs \cite{RuizSimo:2016ikw},
and finite state nucleons are assigned momenta using the phase space model \cite{Sobczyk:2012ms}. For argon an effective model accounting for isospin asymmetry is used \cite{Schwehr:2016pvn}. NEUT partially accounts for a lack of MEC with a large effective axial mass in CCQE events $M_A=1.21$~GeV. Also, NEUT assumes 
that 20\% of $\Delta$'s decay without producing a pion.

There are also some differences in the final state interactions models. For pions NEUT and NuWro use Oset model \cite{Salcedo:1987md}, which in the case of NEUT was futher fine tuned to pion-carbon cross section data. GENIE uses an effective model assuming pion absorption cross section to be a fixed fraction of the pion reaction cross section.

\section{Results}
\label{sec:results}

In this section we discuss the performance of the proposed method to select a high CCQE purity sample of CCQE-like events. We did many tests with a variety of fluxes, targets
and MC event generators. In all the examples we investigated the conclusions were similar.

\begin{table}[h!]
  \begin{tabular}{c|c|c|c|c|c|c}
  Target &Flux & MC & model & threshold  & overall  & CCQE \\
  &&&&[MeV/c]& fraction& purity \\
  \hline
  $^{12}C$ & NuWro &MB & SF & 400& 36.3\% & 82.4\% \\
$^{12}C$ & NuWro  &MB& LFG & 400 & 40.5\%  & 85.0\%  \\
$^{12}C$  & NEUT  &MB& FG &400 & 41.5\% & 90.9\% \\
$^{12}C$ & GENIE  &MB& FG &400& 30.4\% & 91.3\% \\
$^{12}C$ & NEUT &T2K & FG &400& 39.8\% & 91.5\%  \\
$^{12}C$ & GENIE &NuMI & FG &400& 10.0\% & 78.3\%  \\
$^{40}Ar$ & NuWro &MB & LFG & 400 & 39.0\% & 80.2\%  \\
$^{40}Ar$ & NuWro  &MB& SF &400 & 35.2\%  & 77.3\% \\
$^{40}Ar$ & GENIE  &MB& FG &400& 29.1\% & 87.6\% \\
$^{40}Ar$ & NuWro  &MB& LFG &200 & 39.0\% & 89.5\% \\
$^{40}Ar$ & NuWro  &MB& SF &200 & 41.4\%  & 88.5\% \\
$^{40}Ar$ & GENIE  &MB& FG &200& 37.3\% & 95.8\% \\
\hline
\end{tabular}
  \caption{Basic information about MC simulations done in this paper. See explanations in the text.
  }
  \label{basic}
\end{table}

Table \ref{basic} shows basic information about all the simulations that were tested.
We select charged current events with no mesons and exactly one proton above the assumed detection threshold with the MiniBooNE, NuMI and T2K $\nu_{\mu}$ fluxes. The targets we consider are carbon and argon. In the case of carbon the proton threshold is taken to be $400$~MeV/c. In the case of argon there is more flexibility on how do we define the one muon and one proton event sample. We considered two options: either proton threshold at $400$~MeV/c to allow for comparisons with carbon, and also $200$~MeV/c which is the lowest threshold one can expect to achieve in experiments like MicroBooNE.
In all cases, protons below threshold are assumed to be invisible.

In the Table \ref{basic} the sixth column shows what fraction of CC events meet these initial CCQE-like criteria.
In the last column the CCQE purity in the sample is shown. Differences seen in the Table \ref{basic} for the same target and flux express a level of uncertainty of MC generators. For example in the case of MB flux and carbon target the overall fraction according to GENIE is only $\sim 30\%$ while according to NEUT and GENIE it is much larger, about $40\%$. 
This is because GENIE predicts a lower average proton momentum and fewer events pass the $400~MeV/c$ threshold. 
As for the CCQE purity NuWro results (lower purity) are probably more reliable because MEC events are included in the simulation. Fermi gas/SF difference can be explained by the fact that SF predicts smaller CCQE cross section \cite{Benhar:2006nr}. For the T2K flux numbers are similar while for the NuMI flux with an average energy of $\sim 3$~GeV there are many more inelastic events, typically with many pions, so the sample of events with no mesons and only one visible proton is a much lower fraction of the events.

As an illustration for the numbers shown in Table \ref{basic} we present in Table \ref{tab:nuwro} more details about NuWro simulations with MB flux and carbon target. It should be stressed that in the case of MEC events the numbers in the last two columns depend on assumptions in the MEC hadronic model that are very uncertain. We did analogous study using NEUT and we got consistent results for the dominant CCQE contribution.

\begin{table}[h!]
  \centering
  \begin{tabular}{l|c|c|c|c}
    Mode & Overall & CCQE-like & a proton & exactly 1 proton\\
     & & & $>400$MeV/c & $>400$MeV/c \\
    \hline
    CCQE & 51.5\% & 50.2\% & 35.7\% & 34.1\%\\
    RES& 34.0\% & 5.0\% & 4.7\% & 3.1\%\\
    MEC& 10.1\% & 9.8\% & 7.7\%  & 3.3\%\\
    &&&&=40.5\%
  \end{tabular}
  \caption{Breakdown of NuWro (LFG) signal events into interaction modes in a simulation done with MB flux and carbon target.}
    \label{tab:nuwro}
\end{table}

\subsection{Reconstructed Initial Neutron Momentum}

\begin{figure}
  \centering
  \includegraphics[width = 0.52\textwidth]{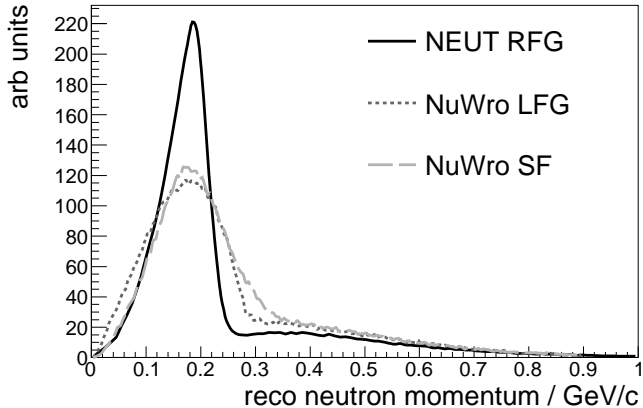}
  \caption{Reconstructed initial state neutron momentum from one muon and one proton events for the global Relativistic Fermi Gas (RFG), Local Fermi Gas (LFG), and Spectral Function (SF) nuclear models. The target nucleus is carbon, and the incident neutrino flux is the MiniBooNE $\nu_{\mu}$ flux.  NEUT is used to produce the RFG simulation, while NuWro is used for LFG and SF samples.
Each sample is normalised to have the same area.}
  \label{fig:neutronMom_nuclearModels}
\end{figure}

\begin{figure}
  \centering
  \includegraphics[width = 0.52\textwidth]{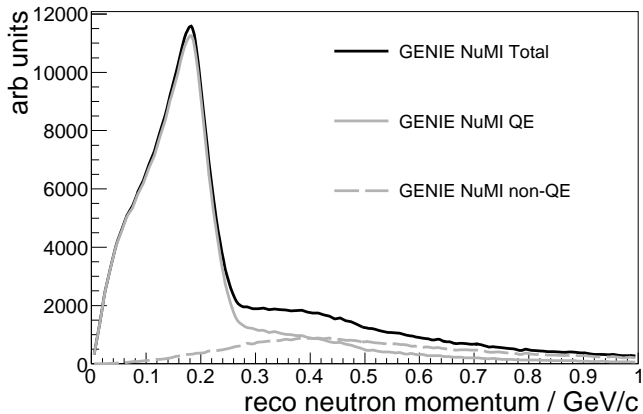}
  \caption{Reconstructed initial state neutron momentum assuming the default GENIE RFG model. The target nucleus is carbon, and the incident neutrino flux is the NuMI on-axis $\nu_{\mu}$ flux. Contributions from CCQE events and non-CCQE events are shown separately.}
  \label{fig:neutronMom_GENIE_NuMI}
\end{figure}

The reconstructed initial state neutron distributions for multiple nuclear models are shown in Fig. \ref{fig:neutronMom_nuclearModels}. We clearly see the typical shape of the nucleon momentum distributions from different models implemented in MCs with a peak at $\sim 200-250$~MeV/c.
In addition there is always a long tail extending to larger values of reconstructed neutron momentum. In Fig. \ref{fig:neutronMom_GENIE_NuMI} reconstructed neutron momentum distribution is shown for the GENIE generator, using the NuMI $\nu_{\mu}$ on-axis flux. Again the target nucleus is carbon.
Here we show contributions from CCQE and non-QE events separately.  It can be clearly seen that the non-QE contribution is largely above the Fermi momentum, while true QE events are usually below the Fermi momentum.
Similar structure was obtained in all the examples we considered. We conclude that imposition of a cut on reconstructed neutron momentum $p_{rec}$ and rejection of events with large values of $p_{rec}$ should select a high purity sample of CCQE events.


\subsection{Cut optimization}

\begin{figure}
  \centering
  \includegraphics[width=0.52\textwidth]{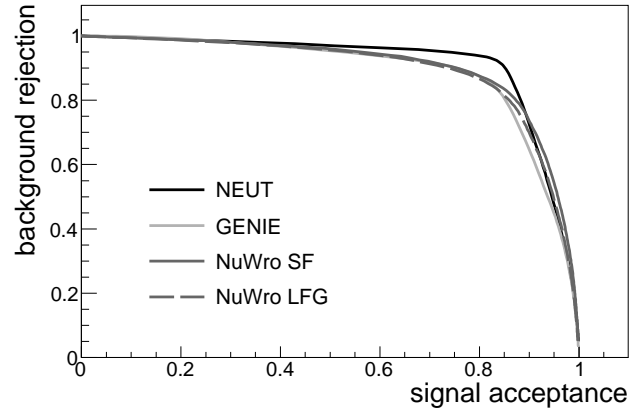}
  \caption{Signal acceptance fraction vs background rejection fraction when cutting on the reconstructed neutron momentum for CCQE-like events.
  Shown are the default configurations of NEUT (solid black) and GENIE (solid light gray), as well as NuWro run with a local Fermi Gas model (solid dark gray) and the Spectral Function model (dashed dark gray).
  In all cases the target nucleus is carbon, and the MiniBooNE $\nu_{\mu}$ flux is used.}
  \label{fig:effRej_generator}
\end{figure}

\begin{figure}
  \centering
  \includegraphics[width=0.52\textwidth]{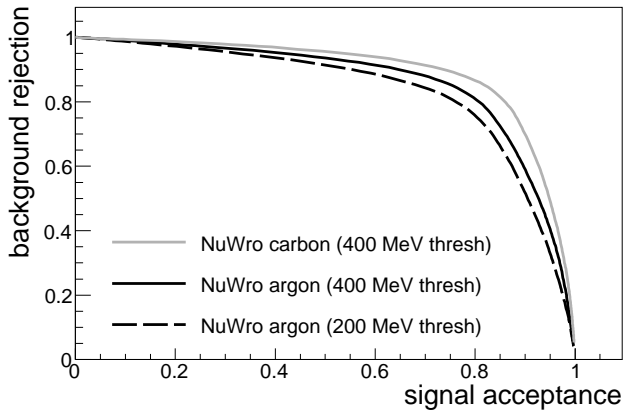}
  \caption{Signal acceptance fraction vs background rejection fraction as a function of a cut on the reconstructed neutron momentum.
  Shown are predictions from NuWro for both carbon (solid gray) and argon (solid black) assuming a 400 MeV/c proton tracking threshold, and for argon also a curve assuming a lower proton tracking threshold of 200 MeV/c (dashed black).}
  \label{fig:effRej_nucleus}
\end{figure}

The level of signal - background separation is found to be very good across all samples tested. Results are shown in Figs \ref{fig:effRej_generator} and \ref{fig:effRej_nucleus}. They present curves of signal (true CCQE) acceptance vs background (true non-CCQE) rejection as a function of a cut on the reconstructed neutron momentum. Each point on the curve corresponds to a value of reconstructed neutron momentum cut.

In Fig. \ref{fig:effRej_nucleus} we can see that the lower threshold degrades performance as nuclear transparency decreases at lower proton momenta. In this case it may be beneficial to artificially impose a higher threshold for protons, though this would also impact the available statistics.

The performance does not vary much as variables such as the generator, target nucleus, and proton tracking threshold, are changed.
Most of the curves allow for over $80\%$ signal acceptance with over $80 \%$ background rejection.

Our studies suggest that a cut at $300$~MeV/c reconstructed neutron momentum provides the best performance in terms of efficiency and purity, though detector resolution effects may slightly modify this conclusion.

\begin{table}[h!]
  \begin{tabular}{c|c|c|c|c|c|c}
  Target & MC & Flux & model & threshold  & overall  & CCQE\\
  &&&&[MeV/c]& fraction& purity\\
  \hline
  $^{12}C$ & NuWro&MB & SF & 400& 26.7\% & 96.0\%\\
$^{12}C$ & NuWro &MB& LFG & 400 & 31.8\%  & 95.5\% \\
$^{12}C$  & NEUT &MB& FG &400 & 33.5\% & 98.2\% \\
$^{12}C$ & GENIE  &MB& FG &400& 24.7\% & 97.4\% \\
$^{12}C$ & NEUT &T2K & FG &400& 32.5\% & 98.5\%  \\
$^{12}C$ & GENIE &NuMI & FG &400& 7.0\% & 95.3\%  \\
$^{40}Ar$ & NuWro &MB& LFG & 400 & 27.3\% & 94.7\% \\
$^{40}Ar$ & NuWro &MB& SF &400 & 23.2\%  & 93.5\%\\
$^{40}Ar$ & GENIE  &MB& FG &400& 21.5\% & 96.3\% \\
$^{40}Ar$ & NuWro &MB& LFG &200 & 33.2\% & 96.6\%\\
$^{40}Ar$ & NuWro &MB& SF &200 & 30.3\%  & 96.3\%\\
$^{40}Ar$ & GENIE  &MB& FG &200& 28.9\% & 98.7\% \\

\hline
\end{tabular}
  \caption{Values of selection efficiency (fraction of true CC events selected) and CCQE purity for different generators, targets, and models after impposing a cut on the reconstructed neutron momentum cut $300$~MeV/c.}
  \label{aftercut}
\end{table}

In Table \ref{aftercut} we show the effect of the cut at 300~MeV/c on the simulated samples of events presented before in Table \ref{basic}. In all the situations one obtains a sample with a purity of $\sim 95$\%, in several cases even better. As we will see in the next section for this sample of events the neutrino energy is reconstructed with a very good precision.


\subsection{Energy Reconstruction}

Often, the neutrino energy is reconstructed using information from final state muon only. Under the assumption that the target neutron is at rest and using energy and momentum conservation (in a similar way to that presented in Sec. \ref{sec:one}) one obtains

\begin{eqnarray}
\label{eq:Eccqe}
E_{CCQE} = \frac{M_p^2 - m^2 +2E'\tilde M_n - \tilde M_n^2}{2(\tilde M_n - E' + k'\cos\theta_\mu)}
\label{eq:ccqeErec}
\end{eqnarray}
where $M_p$, $M_n$ are the proton and neutron masses, $\tilde M_n=M_n-B$ with $B$ binding energy, $m$ is the muon mass, $\theta_\mu$ is the angle the muon makes with the neutrino three-momentum.

When the hadronic final state can be reconstructed, neutrino energy reconstruction usually relies on the hadronic energy deposited in the detector - the exact details depend on the detector technology.
In this case corrections must be made according to MC predictions for the amount of energy carried away by neutrons and the residual nucleus, or particles below threshold.

For events with the topology identified here, a cut on the neutron momentum has been shown to select a high purity sample of CCQE events which are either free from FSI or where FSI effects are mild. For these events, the neutrino energy can be accurately reconstructed with far less concern for missing energy.

\begin{figure}
  \centering
  \includegraphics[width=0.52\textwidth]{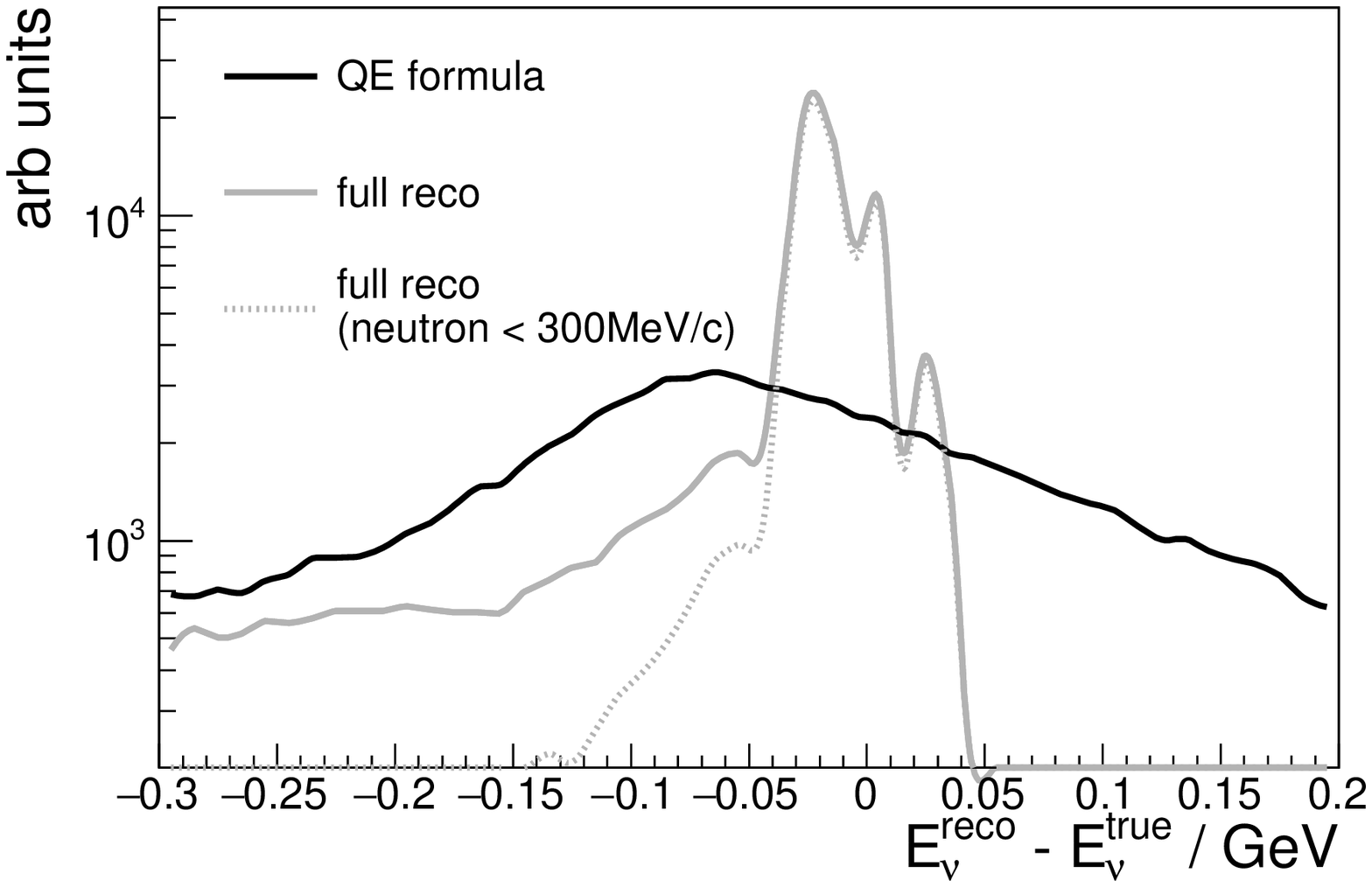}
  \includegraphics[width=0.52\textwidth]{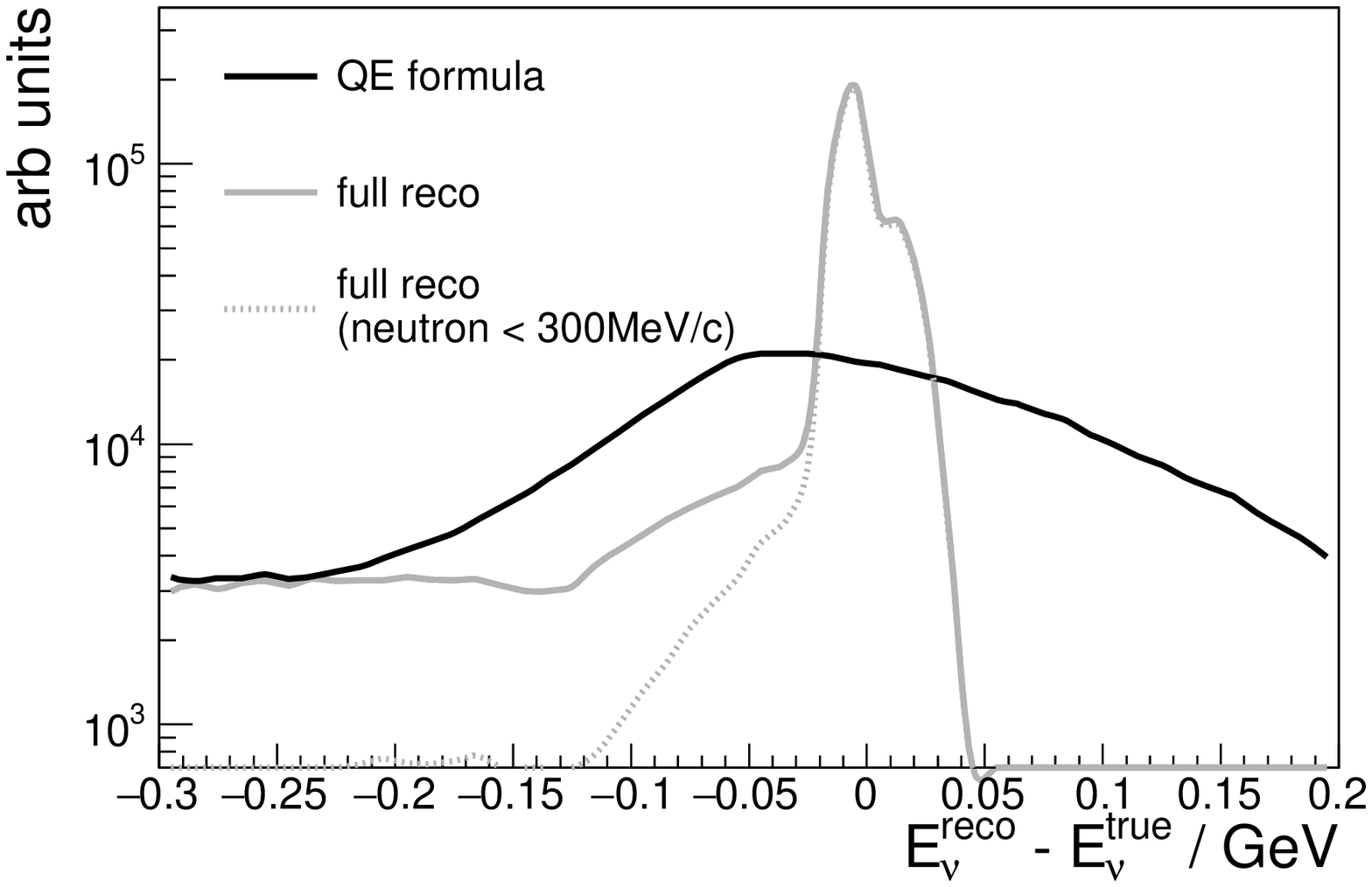}
  \label{fig:Eres}
  \caption{Neutrino energy resolution for $1\mu\ 1p$ events, using the CCQE formula (black solid), the method proposed in this paper without (gray solid), and with a cut on the reconstructed neutron momentum at 300MeV/c (gray dotted). 
  The target nucleus used is argon (top) or carbon (bottom) and the beam is MiniBooNE's $\nu_\mu$ (top) and NuMI's on-axis $\nu_{\mu}$ (bottom).
  }
\end{figure}

Fig. \ref{fig:Eres} shows the difference between the true and reconstructed neutrino energy  for a sample of CCQE-like $1\mu\ 1p$ events discussed in this paper, using our method (Eq. \ref{eq:Eccqe}) with and without a cut on the reconstructed neutron momentum of 300 MeV/c. The cut removes most of events with poorly reconstructed neutrino energy, leaving a sample of events with neutrino energies reconstructed within 100 MeV of the true neutrino energy with a vast majority of events reconstructed much better. Characteristic shapes seen in Fig. \ref{fig:Eres} comes from the probability distribution for neutron binding energy. They are present because in the simulations Fermi gas model was used with no information about neutron energy levels. Of course, in a real experimental situation the structures will likely be washed out by detector smearing effects.

It is interesting to observe that FSI tends to lower energy of final state particles, leading to appearance of a tail of events being reconstructed with the energy often much smaller than the true one.  Placing a cut at $300$~MeV/c reconstructed neutron momentum reduces this tail significantly.

Solid line denotes a performance of the energy reconstruction based on Eq. (\ref{eq:ccqeErec}). The method proposed in this paper is dramatically better.

\section{Discussion}
\label{sec:discussion}

In the case of argon target with a very low proton detection threshold there are many strategies in which one can get even better CCQE purity. For example, one can select events with only one proton above $200$~MeV/c and also only one proton with momentum above $400$~MeV/c. This is a more restrictive sample than discussed before as there are events with one proton above 400~MeV/c and the subleading one between 200 and 400~MeV/c.

According to NuWro run with  SF the above 
sample of events contains $\sim 27.7$\% of the overall number of CC events. CCQE purity of this sample is $88.4$\%. A cut of the reconstructed neutron momentum at $300$~MeV/c reduces this sample to  $\sim 21.3$\% of CC events with the purity as high as $\sim 97.1$\%. According to simulations done with LFG the purity of the selected sample is even better: $\sim 97.7$\%.

An interesting question is if the cut discussed in this paper can tell us about RES and MEC events separately. We looked at the distributions of neutron momentum resulting from RES and MEC events in NuWro simulations with MB beam on the carbon target. RES events give rise to very flat distribution of reconstructed momenta while MEC events show more structure: a gentle maximum at $\sim 450$~MeV/c with a wide spread. However, in both cases the results depend on several models with large uncertainties and we think it may be risky to try to use this technique to get information about a size of the MEC contribution.

In a real experimental set-up the target is never pure carbon but usually CH or CH$_2$. In such a case our conclusion about the CCQE purity of the selected sample remains valid as neutrino interactions on hydrogen gives no contribution to CCQE-like sample. The only difference is in the relative normalization of the selected sample. It is smaller and the size of the difference will depend on neutrino flux spectrum.

\section{Conclusions}
\label{sec:conclusions}

We proposed a method to select a high purity CCQE sample of events. We checked the performance of the method using different neutrino fluxes, targets and MC generators. In all the cases the conclusion is that the purity of the sample is $\sim 95\%$. We also checked that in the selected sample neutrino energy is reconstructed with a very good precision.

\begin{acknowledgments}
JTS was supported by Fermilab Neutrino Physics Center Fellowship and also by NCN grant UMO-2014/14/M/ST2/00850.
APF was supported by the UK Science and Technology Funding Council, and also by the Fermilab Intensity Frontier Fellowship.
\end{acknowledgments}

\bibliographystyle{apsrev4-1}

\bibliography{bibliography}

\begin{thebibliography}{40}%
\makeatletter
\providecommand \@ifxundefined [1]{%
 \@ifx{#1\undefined}
}%
\providecommand \@ifnum [1]{%
 \ifnum #1\expandafter \@firstoftwo
 \else \expandafter \@secondoftwo
 \fi
}%
\providecommand \@ifx [1]{%
 \ifx #1\expandafter \@firstoftwo
 \else \expandafter \@secondoftwo
 \fi
}%
\providecommand \natexlab [1]{#1}%
\providecommand \enquote  [1]{``#1''}%
\providecommand \bibnamefont  [1]{#1}%
\providecommand \bibfnamefont [1]{#1}%
\providecommand \citenamefont [1]{#1}%
\providecommand \href@noop [0]{\@secondoftwo}%
\providecommand \href [0]{\begingroup \@sanitize@url \@href}%
\providecommand \@href[1]{\@@startlink{#1}\@@href}%
\providecommand \@@href[1]{\endgroup#1\@@endlink}%
\providecommand \@sanitize@url [0]{\catcode `\\12\catcode `\$12\catcode
  `\&12\catcode `\#12\catcode `\^12\catcode `\_12\catcode `\%12\relax}%
\providecommand \@@startlink[1]{}%
\providecommand \@@endlink[0]{}%
\providecommand \url  [0]{\begingroup\@sanitize@url \@url }%
\providecommand \@url [1]{\endgroup\@href {#1}{\urlprefix }}%
\providecommand \urlprefix  [0]{URL }%
\providecommand \Eprint [0]{\href }%
\providecommand \doibase [0]{http://dx.doi.org/}%
\providecommand \selectlanguage [0]{\@gobble}%
\providecommand \bibinfo  [0]{\@secondoftwo}%
\providecommand \bibfield  [0]{\@secondoftwo}%
\providecommand \translation [1]{[#1]}%
\providecommand \BibitemOpen [0]{}%
\providecommand \bibitemStop [0]{}%
\providecommand \bibitemNoStop [0]{.\EOS\space}%
\providecommand \EOS [0]{\spacefactor3000\relax}%
\providecommand \BibitemShut  [1]{\csname bibitem#1\endcsname}%
\let\auto@bib@innerbib\@empty
\bibitem [{\citenamefont {Aguilar-Arevalo~{\it et al.}
  [MiniBooNE~Collaboration]}(2010)}]{AguilarArevalo:2010zc}%
  \BibitemOpen
  \bibfield  {author} {\bibinfo {author} {\bibfnamefont {A.~A.}\ \bibnamefont
  {Aguilar-Arevalo~{\it et al.} [MiniBooNE~Collaboration]}},\ }\href@noop {}
  {\bibfield  {journal} {\bibinfo  {journal} {Phys.\ Rev.\ D}\ }\textbf
  {\bibinfo {volume} {{\bf 81}}},\ \bibinfo {pages} {092005} (\bibinfo {year}
  {2010})}\BibitemShut {NoStop}%
\bibitem [{\citenamefont {Martini}\ \emph {et~al.}(2009)\citenamefont
  {Martini}, \citenamefont {Ericson}, \citenamefont {Chanfray},\ and\
  \citenamefont {Marteau}}]{Martini:2009uj}%
  \BibitemOpen
  \bibfield  {author} {\bibinfo {author} {\bibfnamefont {M.}~\bibnamefont
  {Martini}}, \bibinfo {author} {\bibfnamefont {M.}~\bibnamefont {Ericson}},
  \bibinfo {author} {\bibfnamefont {G.}~\bibnamefont {Chanfray}}, \ and\
  \bibinfo {author} {\bibfnamefont {J.}~\bibnamefont {Marteau}},\ }\href@noop
  {} {\bibfield  {journal} {\bibinfo  {journal} {Phys.\ Rev.\ C}\ }\textbf
  {\bibinfo {volume} {{\bf 80}}},\ \bibinfo {pages} {065501} (\bibinfo {year}
  {2009})}\BibitemShut {NoStop}%
\bibitem [{\citenamefont {Nieves}\ \emph {et~al.}(2011)\citenamefont {Nieves},
  \citenamefont {Ruiz~Simo},\ and\ \citenamefont
  {Vicente~Vacas}}]{Nieves:2011pp}%
  \BibitemOpen
  \bibfield  {author} {\bibinfo {author} {\bibfnamefont {J.}~\bibnamefont
  {Nieves}}, \bibinfo {author} {\bibfnamefont {I.}~\bibnamefont {Ruiz~Simo}}, \
  and\ \bibinfo {author} {\bibfnamefont {M.~J.}\ \bibnamefont
  {Vicente~Vacas}},\ }\href@noop {} {\bibfield  {journal} {\bibinfo  {journal}
  {Phys.\ Rev.\ C}\ }\textbf {\bibinfo {volume} {83}},\ \bibinfo {pages}
  {045501} (\bibinfo {year} {2011})}\BibitemShut {NoStop}%
\bibitem [{\citenamefont {Martini}\ \emph {et~al.}(2012)\citenamefont
  {Martini}, \citenamefont {Ericson},\ and\ \citenamefont
  {Chanfray}}]{Martini:2012fa}%
  \BibitemOpen
  \bibfield  {author} {\bibinfo {author} {\bibfnamefont {M.}~\bibnamefont
  {Martini}}, \bibinfo {author} {\bibfnamefont {M.}~\bibnamefont {Ericson}}, \
  and\ \bibinfo {author} {\bibfnamefont {G.}~\bibnamefont {Chanfray}},\
  }\href@noop {} {\bibfield  {journal} {\bibinfo  {journal} {Phys.\ Rev.\ D}\
  }\textbf {\bibinfo {volume} {85}},\ \bibinfo {pages} {093012} (\bibinfo
  {year} {2012})}\BibitemShut {NoStop}%
\bibitem [{\citenamefont {Lalakulich}\ \emph {et~al.}(2012)\citenamefont
  {Lalakulich}, \citenamefont {Mosel},\ and\ \citenamefont
  {Gallmeister}}]{Lalakulich:2012hs}%
  \BibitemOpen
  \bibfield  {author} {\bibinfo {author} {\bibfnamefont {O.}~\bibnamefont
  {Lalakulich}}, \bibinfo {author} {\bibfnamefont {U.}~\bibnamefont {Mosel}}, \
  and\ \bibinfo {author} {\bibfnamefont {K.}~\bibnamefont {Gallmeister}},\
  }\href {\doibase 10.1103/PhysRevC.86.054606} {\bibfield  {journal} {\bibinfo
  {journal} {Phys. Rev.}\ }\textbf {\bibinfo {volume} {C86}},\ \bibinfo {pages}
  {054606} (\bibinfo {year} {2012})},\ \Eprint {http://arxiv.org/abs/1208.3678}
  {arXiv:1208.3678 [nucl-th]} \BibitemShut {NoStop}%
\bibitem [{\citenamefont {Nieves}\ \emph {et~al.}(2012)\citenamefont {Nieves},
  \citenamefont {Sanchez}, \citenamefont {Ruiz~Simo},\ and\ \citenamefont
  {Vicente~Vacas}}]{Nieves:2012yz}%
  \BibitemOpen
  \bibfield  {author} {\bibinfo {author} {\bibfnamefont {J.}~\bibnamefont
  {Nieves}}, \bibinfo {author} {\bibfnamefont {F.}~\bibnamefont {Sanchez}},
  \bibinfo {author} {\bibfnamefont {I.}~\bibnamefont {Ruiz~Simo}}, \ and\
  \bibinfo {author} {\bibfnamefont {M.~J.}\ \bibnamefont {Vicente~Vacas}},\
  }\href {\doibase 10.1103/PhysRevD.85.113008} {\bibfield  {journal} {\bibinfo
  {journal} {Phys. Rev.}\ }\textbf {\bibinfo {volume} {D85}},\ \bibinfo {pages}
  {113008} (\bibinfo {year} {2012})},\ \Eprint {http://arxiv.org/abs/1204.5404}
  {arXiv:1204.5404 [hep-ph]} \BibitemShut {NoStop}%
\bibitem [{\citenamefont {Morfin}\ \emph {et~al.}(2012)\citenamefont {Morfin},
  \citenamefont {Nieves},\ and\ \citenamefont {Sobczyk}}]{Morfin:2012kn}%
  \BibitemOpen
  \bibfield  {author} {\bibinfo {author} {\bibfnamefont {J.~G.}\ \bibnamefont
  {Morfin}}, \bibinfo {author} {\bibfnamefont {J.}~\bibnamefont {Nieves}}, \
  and\ \bibinfo {author} {\bibfnamefont {J.~T.}\ \bibnamefont {Sobczyk}},\
  }\href@noop {} {\bibfield  {journal} {\bibinfo  {journal} {Adv.\ High Energy
  Phys.}\ }\textbf {\bibinfo {volume} {2012}},\ \bibinfo {pages} {934597}
  (\bibinfo {year} {2012})}\BibitemShut {NoStop}%
\bibitem [{\citenamefont {Alvarez-Ruso}\ \emph {et~al.}(2014)\citenamefont
  {Alvarez-Ruso}, \citenamefont {Hayato},\ and\ \citenamefont
  {Nieves}}]{Alvarez-Ruso:2014bla}%
  \BibitemOpen
  \bibfield  {author} {\bibinfo {author} {\bibfnamefont {L.}~\bibnamefont
  {Alvarez-Ruso}}, \bibinfo {author} {\bibfnamefont {Y.}~\bibnamefont
  {Hayato}}, \ and\ \bibinfo {author} {\bibfnamefont {J.}~\bibnamefont
  {Nieves}},\ }\href {\doibase 10.1088/1367-2630/16/7/075015} {\bibfield
  {journal} {\bibinfo  {journal} {New J. Phys.}\ }\textbf {\bibinfo {volume}
  {16}},\ \bibinfo {pages} {075015} (\bibinfo {year} {2014})},\ \Eprint
  {http://arxiv.org/abs/1403.2673} {arXiv:1403.2673 [hep-ph]} \BibitemShut
  {NoStop}%
\bibitem [{\citenamefont {Benhar}\ \emph {et~al.}(2005)\citenamefont {Benhar},
  \citenamefont {Farina}, \citenamefont {Nakamura}, \citenamefont {Sakuda},\
  and\ \citenamefont {Seki}}]{Benhar:2005dj}%
  \BibitemOpen
  \bibfield  {author} {\bibinfo {author} {\bibfnamefont {O.}~\bibnamefont
  {Benhar}}, \bibinfo {author} {\bibfnamefont {N.}~\bibnamefont {Farina}},
  \bibinfo {author} {\bibfnamefont {H.}~\bibnamefont {Nakamura}}, \bibinfo
  {author} {\bibfnamefont {M.}~\bibnamefont {Sakuda}}, \ and\ \bibinfo {author}
  {\bibfnamefont {R.}~\bibnamefont {Seki}},\ }\href@noop {} {\bibfield
  {journal} {\bibinfo  {journal} {Phys.\ Rev.\ D}\ }\textbf {\bibinfo {volume}
  {72}},\ \bibinfo {pages} {053005} (\bibinfo {year} {2005})}\BibitemShut
  {NoStop}%
\bibitem [{\citenamefont {Rodrigues}\ \emph {et~al.}(2016)\citenamefont
  {Rodrigues} \emph {et~al.}}]{Rodrigues:2015hik}%
  \BibitemOpen
  \bibfield  {author} {\bibinfo {author} {\bibfnamefont {P.~A.}\ \bibnamefont
  {Rodrigues}} \emph {et~al.} (\bibinfo {collaboration} {MINERvA}),\ }\href
  {\doibase 10.1103/PhysRevLett.116.071802} {\bibfield  {journal} {\bibinfo
  {journal} {Phys. Rev. Lett.}\ }\textbf {\bibinfo {volume} {116}},\ \bibinfo
  {pages} {071802} (\bibinfo {year} {2016})},\ \Eprint
  {http://arxiv.org/abs/1511.05944} {arXiv:1511.05944 [hep-ex]} \BibitemShut
  {NoStop}%
\bibitem [{\citenamefont {Abe}\ \emph {et~al.}(2016)\citenamefont {Abe} \emph
  {et~al.}}]{Abe:2016tmq}%
  \BibitemOpen
  \bibfield  {author} {\bibinfo {author} {\bibfnamefont {K.}~\bibnamefont
  {Abe}} \emph {et~al.} (\bibinfo {collaboration} {T2K}),\ }\href {\doibase
  10.1103/PhysRevD.93.112012} {\bibfield  {journal} {\bibinfo  {journal} {Phys.
  Rev.}\ }\textbf {\bibinfo {volume} {D93}},\ \bibinfo {pages} {112012}
  (\bibinfo {year} {2016})},\ \Eprint {http://arxiv.org/abs/1602.03652}
  {arXiv:1602.03652 [hep-ex]} \BibitemShut {NoStop}%
\bibitem [{\citenamefont {Gallagher}\ and\ \citenamefont
  {Hayato}(2014)}]{PDG-MC}%
  \BibitemOpen
  \bibfield  {author} {\bibinfo {author} {\bibfnamefont {H.}~\bibnamefont
  {Gallagher}}\ and\ \bibinfo {author} {\bibfnamefont {Y.}~\bibnamefont
  {Hayato}} (\bibinfo {collaboration} {Particle Data Group}),\ }\href {\doibase
  10.1088/1674-1137/38/9/090001} {\bibfield  {journal} {\bibinfo  {journal}
  {Chin. Phys.}\ }\textbf {\bibinfo {volume} {C38}},\ \bibinfo {pages} {090001}
  (\bibinfo {year} {2014})}\BibitemShut {NoStop}%
\bibitem [{\citenamefont {Walton}\ \emph {et~al.}(2015)\citenamefont {Walton}
  \emph {et~al.}}]{Walton:2014esl}%
  \BibitemOpen
  \bibfield  {author} {\bibinfo {author} {\bibfnamefont {T.}~\bibnamefont
  {Walton}} \emph {et~al.} (\bibinfo {collaboration} {MINERvA}),\ }\href
  {\doibase 10.1103/PhysRevD.91.071301} {\bibfield  {journal} {\bibinfo
  {journal} {Phys. Rev.}\ }\textbf {\bibinfo {volume} {D91}},\ \bibinfo {pages}
  {071301} (\bibinfo {year} {2015})},\ \Eprint {http://arxiv.org/abs/1409.4497}
  {arXiv:1409.4497 [hep-ex]} \BibitemShut {NoStop}%
\bibitem [{\citenamefont {Garino}\ \emph {et~al.}(1992)\citenamefont {Garino}
  \emph {et~al.}}]{Garino:1992ca}%
  \BibitemOpen
  \bibfield  {author} {\bibinfo {author} {\bibfnamefont {G.}~\bibnamefont
  {Garino}} \emph {et~al.},\ }\href {\doibase 10.1103/PhysRevC.45.780}
  {\bibfield  {journal} {\bibinfo  {journal} {Phys. Rev.}\ }\textbf {\bibinfo
  {volume} {C45}},\ \bibinfo {pages} {780} (\bibinfo {year}
  {1992})}\BibitemShut {NoStop}%
\bibitem [{\citenamefont {O'Neill}\ \emph {et~al.}(1995)\citenamefont {O'Neill}
  \emph {et~al.}}]{O'Neill:1994mg}%
  \BibitemOpen
  \bibfield  {author} {\bibinfo {author} {\bibfnamefont {T.~G.}\ \bibnamefont
  {O'Neill}} \emph {et~al.},\ }\href {\doibase 10.1016/0370-2693(95)00362-O}
  {\bibfield  {journal} {\bibinfo  {journal} {Phys. Lett.}\ }\textbf {\bibinfo
  {volume} {B351}},\ \bibinfo {pages} {87} (\bibinfo {year} {1995})},\ \Eprint
  {http://arxiv.org/abs/hep-ph/9408260} {arXiv:hep-ph/9408260 [hep-ph]}
  \BibitemShut {NoStop}%
\bibitem [{\citenamefont {Hen}\ \emph {et~al.}(2013)\citenamefont {Hen} \emph
  {et~al.}}]{Hen:2012yva}%
  \BibitemOpen
  \bibfield  {author} {\bibinfo {author} {\bibfnamefont {O.}~\bibnamefont
  {Hen}} \emph {et~al.} (\bibinfo {collaboration} {CLAS}),\ }\href {\doibase
  10.1016/j.physletb.2013.04.011} {\bibfield  {journal} {\bibinfo  {journal}
  {Phys. Lett.}\ }\textbf {\bibinfo {volume} {B722}},\ \bibinfo {pages} {63}
  (\bibinfo {year} {2013})},\ \Eprint {http://arxiv.org/abs/1212.5343}
  {arXiv:1212.5343 [nucl-ex]} \BibitemShut {NoStop}%
\bibitem [{\citenamefont {Lu}\ \emph {et~al.}(2016)\citenamefont {Lu},
  \citenamefont {Pickering}, \citenamefont {Dolan}, \citenamefont {Barr},
  \citenamefont {Coplowe}, \citenamefont {Uchida}, \citenamefont {Wark},
  \citenamefont {Wascko}, \citenamefont {Weber},\ and\ \citenamefont
  {Yuan}}]{Lu:2015tcr}%
  \BibitemOpen
  \bibfield  {author} {\bibinfo {author} {\bibfnamefont {X.~G.}\ \bibnamefont
  {Lu}}, \bibinfo {author} {\bibfnamefont {L.}~\bibnamefont {Pickering}},
  \bibinfo {author} {\bibfnamefont {S.}~\bibnamefont {Dolan}}, \bibinfo
  {author} {\bibfnamefont {G.}~\bibnamefont {Barr}}, \bibinfo {author}
  {\bibfnamefont {D.}~\bibnamefont {Coplowe}}, \bibinfo {author} {\bibfnamefont
  {Y.}~\bibnamefont {Uchida}}, \bibinfo {author} {\bibfnamefont
  {D.}~\bibnamefont {Wark}}, \bibinfo {author} {\bibfnamefont {M.~O.}\
  \bibnamefont {Wascko}}, \bibinfo {author} {\bibfnamefont {A.}~\bibnamefont
  {Weber}}, \ and\ \bibinfo {author} {\bibfnamefont {T.}~\bibnamefont {Yuan}},\
  }\href {\doibase 10.1103/PhysRevC.94.015503} {\bibfield  {journal} {\bibinfo
  {journal} {Phys. Rev.}\ }\textbf {\bibinfo {volume} {C94}},\ \bibinfo {pages}
  {015503} (\bibinfo {year} {2016})},\ \Eprint
  {http://arxiv.org/abs/1512.05748} {arXiv:1512.05748 [nucl-th]} \BibitemShut
  {NoStop}%
\bibitem [{\citenamefont {Lu}\ and\ \citenamefont
  {Betancourt}(2016)}]{Lu:2016mjf}%
  \BibitemOpen
  \bibfield  {author} {\bibinfo {author} {\bibfnamefont {X.~G.}\ \bibnamefont
  {Lu}}\ and\ \bibinfo {author} {\bibfnamefont {M.}~\bibnamefont {Betancourt}}
  (\bibinfo {collaboration} {MINERvA}),\ }in\ \href
  {https://inspirehep.net/record/1481397/files/arXiv:1608.04655.pdf} {\emph
  {\bibinfo {booktitle} {{27th International Conference on Neutrino Physics and
  Astrophysics (Neutrino 2016) (Neutrino 2016) London, UK, July 4-9, 2016}}}}\
  (\bibinfo {year} {2016})\ \Eprint {http://arxiv.org/abs/1608.04655}
  {arXiv:1608.04655 [hep-ex]} \BibitemShut {NoStop}%
\bibitem [{\citenamefont {Benhar}\ \emph {et~al.}(1994)\citenamefont {Benhar},
  \citenamefont {Fabrocini}, \citenamefont {Fantoni},\ and\ \citenamefont
  {Sick}}]{Benhar:1994hw}%
  \BibitemOpen
  \bibfield  {author} {\bibinfo {author} {\bibfnamefont {O.}~\bibnamefont
  {Benhar}}, \bibinfo {author} {\bibfnamefont {A.}~\bibnamefont {Fabrocini}},
  \bibinfo {author} {\bibfnamefont {S.}~\bibnamefont {Fantoni}}, \ and\
  \bibinfo {author} {\bibfnamefont {I.}~\bibnamefont {Sick}},\ }\href@noop {}
  {\bibfield  {journal} {\bibinfo  {journal} {Nucl.\ Phys.\ A}\ }\textbf
  {\bibinfo {volume} {579}},\ \bibinfo {pages} {493} (\bibinfo {year}
  {1994})}\BibitemShut {NoStop}%
\bibitem [{\citenamefont {Hen}\ \emph {et~al.}(2014)\citenamefont {Hen} \emph
  {et~al.}}]{Hen:2014nza}%
  \BibitemOpen
  \bibfield  {author} {\bibinfo {author} {\bibfnamefont {O.}~\bibnamefont
  {Hen}} \emph {et~al.},\ }\href {\doibase 10.1126/science.1256785} {\bibfield
  {journal} {\bibinfo  {journal} {Science}\ }\textbf {\bibinfo {volume}
  {346}},\ \bibinfo {pages} {614} (\bibinfo {year} {2014})},\ \Eprint
  {http://arxiv.org/abs/1412.0138} {arXiv:1412.0138 [nucl-ex]} \BibitemShut
  {NoStop}%
\bibitem [{\citenamefont {Andreopoulos}\ \emph {et~al.}(2010)\citenamefont
  {Andreopoulos}, \citenamefont {Bell}, \citenamefont {Bhattacharya},
  \citenamefont {Cavanna}, \citenamefont {Dobson} \emph
  {et~al.}}]{Andreopoulos:2009rq}%
  \BibitemOpen
  \bibfield  {author} {\bibinfo {author} {\bibfnamefont {C.}~\bibnamefont
  {Andreopoulos}}, \bibinfo {author} {\bibfnamefont {A.}~\bibnamefont {Bell}},
  \bibinfo {author} {\bibfnamefont {D.}~\bibnamefont {Bhattacharya}}, \bibinfo
  {author} {\bibfnamefont {F.}~\bibnamefont {Cavanna}}, \bibinfo {author}
  {\bibfnamefont {J.}~\bibnamefont {Dobson}},  \emph {et~al.},\ }\href
  {\doibase 10.1016/j.nima.2009.12.009} {\bibfield  {journal} {\bibinfo
  {journal} {Nucl. Instrum. Meth.}\ }\textbf {\bibinfo {volume} {A614}},\
  \bibinfo {pages} {87} (\bibinfo {year} {2010})},\ \Eprint
  {http://arxiv.org/abs/0905.2517} {arXiv:0905.2517 [hep-ph]} \BibitemShut
  {NoStop}%
\bibitem [{\citenamefont {Hayato}(2009)}]{Hayato:2009zz}%
  \BibitemOpen
  \bibfield  {author} {\bibinfo {author} {\bibfnamefont {Y.}~\bibnamefont
  {Hayato}},\ }\bibfield  {booktitle} {\emph {\bibinfo {booktitle} {{Neutrino
  interactions: From theory to Monte Carlo simulations. Proceedings, 45th
  Karpacz Winter School in Theoretical Physics, Ladek-Zdroj, Poland, February
  2-11, 2009}}},\ }\href@noop {} {\bibfield  {journal} {\bibinfo  {journal}
  {Acta Phys. Polon.}\ }\textbf {\bibinfo {volume} {B40}},\ \bibinfo {pages}
  {2477} (\bibinfo {year} {2009})}\BibitemShut {NoStop}%
\bibitem [{\citenamefont {Golan}\ \emph {et~al.}(2012)\citenamefont {Golan},
  \citenamefont {Juszczak},\ and\ \citenamefont {Sobczyk}}]{Golan:2012wx}%
  \BibitemOpen
  \bibfield  {author} {\bibinfo {author} {\bibfnamefont {T.}~\bibnamefont
  {Golan}}, \bibinfo {author} {\bibfnamefont {C.}~\bibnamefont {Juszczak}}, \
  and\ \bibinfo {author} {\bibfnamefont {J.~T.}\ \bibnamefont {Sobczyk}},\
  }\href@noop {} {\bibfield  {journal} {\bibinfo  {journal} {Phys.\ Rev.\ C}\
  }\textbf {\bibinfo {volume} {86}},\ \bibinfo {pages} {015505} (\bibinfo
  {year} {2012})}\BibitemShut {NoStop}%
\bibitem [{\citenamefont {Ankowski}\ and\ \citenamefont
  {Sobczyk}(2008)}]{Ankowski:2007uy}%
  \BibitemOpen
  \bibfield  {author} {\bibinfo {author} {\bibfnamefont {A.~M.}\ \bibnamefont
  {Ankowski}}\ and\ \bibinfo {author} {\bibfnamefont {J.~T.}\ \bibnamefont
  {Sobczyk}},\ }\href@noop {} {\bibfield  {journal} {\bibinfo  {journal}
  {Phys.\ Rev.\ C}\ }\textbf {\bibinfo {volume} {77}},\ \bibinfo {pages}
  {044311} (\bibinfo {year} {2008})}\BibitemShut {NoStop}%
\bibitem [{\citenamefont {Frullani}\ and\ \citenamefont
  {Mougey}(1984)}]{Frullani:1984nn}%
  \BibitemOpen
  \bibfield  {author} {\bibinfo {author} {\bibfnamefont {S.}~\bibnamefont
  {Frullani}}\ and\ \bibinfo {author} {\bibfnamefont {J.}~\bibnamefont
  {Mougey}},\ }\href@noop {} {\bibfield  {journal} {\bibinfo  {journal} {Adv.
  Nucl. Phys.}\ }\textbf {\bibinfo {volume} {14}},\ \bibinfo {pages} {1}
  (\bibinfo {year} {1984})}\BibitemShut {NoStop}%
\bibitem [{\citenamefont {Ankowski}(2012)}]{Ankowski:2012ei}%
  \BibitemOpen
  \bibfield  {author} {\bibinfo {author} {\bibfnamefont {A.~M.}\ \bibnamefont
  {Ankowski}},\ }\href {\doibase 10.1103/PhysRevC.86.024616} {\bibfield
  {journal} {\bibinfo  {journal} {Phys. Rev.}\ }\textbf {\bibinfo {volume}
  {C86}},\ \bibinfo {pages} {024616} (\bibinfo {year} {2012})},\ \Eprint
  {http://arxiv.org/abs/1205.4804} {arXiv:1205.4804 [nucl-th]} \BibitemShut
  {NoStop}%
\bibitem [{\citenamefont {Bodek}\ and\ \citenamefont
  {Ritchie}(1981)}]{Bodek:1980ar}%
  \BibitemOpen
  \bibfield  {author} {\bibinfo {author} {\bibfnamefont {A.}~\bibnamefont
  {Bodek}}\ and\ \bibinfo {author} {\bibfnamefont {J.~L.}\ \bibnamefont
  {Ritchie}},\ }\href {\doibase 10.1103/PhysRevD.23.1070} {\bibfield  {journal}
  {\bibinfo  {journal} {Phys. Rev.}\ }\textbf {\bibinfo {volume} {D23}},\
  \bibinfo {pages} {1070} (\bibinfo {year} {1981})}\BibitemShut {NoStop}%
\bibitem [{\citenamefont {Benhar}\ \emph {et~al.}(1989)\citenamefont {Benhar},
  \citenamefont {Fabrocini},\ and\ \citenamefont {Fantoni}}]{Benhar:1989aw}%
  \BibitemOpen
  \bibfield  {author} {\bibinfo {author} {\bibfnamefont {O.}~\bibnamefont
  {Benhar}}, \bibinfo {author} {\bibfnamefont {A.}~\bibnamefont {Fabrocini}}, \
  and\ \bibinfo {author} {\bibfnamefont {S.}~\bibnamefont {Fantoni}},\
  }\href@noop {} {\bibfield  {journal} {\bibinfo  {journal} {Nucl.\ Phys.\ A}\
  }\textbf {\bibinfo {volume} {505}},\ \bibinfo {pages} {267} (\bibinfo {year}
  {1989})}\BibitemShut {NoStop}%
\bibitem [{\citenamefont {Rein}\ and\ \citenamefont
  {Sehgal}(1981)}]{Rein:1980wg}%
  \BibitemOpen
  \bibfield  {author} {\bibinfo {author} {\bibfnamefont {D.}~\bibnamefont
  {Rein}}\ and\ \bibinfo {author} {\bibfnamefont {L.~M.}\ \bibnamefont
  {Sehgal}},\ }\href@noop {} {\bibfield  {journal} {\bibinfo  {journal} {Annals
  Phys.}\ }\textbf {\bibinfo {volume} {133}},\ \bibinfo {pages} {79} (\bibinfo
  {year} {1981})}\BibitemShut {NoStop}%
\bibitem [{\citenamefont {Radecky}\ \emph {et~al.}(1982)\citenamefont
  {Radecky}, \citenamefont {Barnes}, \citenamefont {Carmony}, \citenamefont
  {Garfinkel}, \citenamefont {Derrick}, \citenamefont {Fernandez},
  \citenamefont {Hyman},\ and\ \citenamefont {Levman~{\it et
  al.}}}]{Radecky:1981fn}%
  \BibitemOpen
  \bibfield  {author} {\bibinfo {author} {\bibfnamefont {G.~M.}\ \bibnamefont
  {Radecky}}, \bibinfo {author} {\bibfnamefont {V.~E.}\ \bibnamefont {Barnes}},
  \bibinfo {author} {\bibfnamefont {D.~D.}\ \bibnamefont {Carmony}}, \bibinfo
  {author} {\bibfnamefont {A.~F.}\ \bibnamefont {Garfinkel}}, \bibinfo {author}
  {\bibfnamefont {M.}~\bibnamefont {Derrick}}, \bibinfo {author} {\bibfnamefont
  {E.}~\bibnamefont {Fernandez}}, \bibinfo {author} {\bibfnamefont
  {L.}~\bibnamefont {Hyman}}, \ and\ \bibinfo {author} {\bibfnamefont
  {G.}~\bibnamefont {Levman~{\it et al.}}},\ }\href@noop {} {\bibfield
  {journal} {\bibinfo  {journal} {Phys.\ Rev.\ D}\ }\textbf {\bibinfo {volume}
  {{\bf 26}}},\ \bibinfo {pages} {3297} (\bibinfo {year} {1982})},\ \bibinfo
  {note} {[Erratum-ibid.\ D {\bf 26} (1982) 3297]}\BibitemShut {NoStop}%
\bibitem [{\citenamefont {Barish}\ \emph {et~al.}(1979)\citenamefont {Barish},
  \citenamefont {Derrick}, \citenamefont {Dombeck}, \citenamefont {Hyman},
  \citenamefont {Jaeger}, \citenamefont {Musgrave}, \citenamefont {Schreiner},\
  and\ \citenamefont {Singer~{\it et al.}}}]{Barish:1978pj}%
  \BibitemOpen
  \bibfield  {author} {\bibinfo {author} {\bibfnamefont {S.~J.}\ \bibnamefont
  {Barish}}, \bibinfo {author} {\bibfnamefont {M.}~\bibnamefont {Derrick}},
  \bibinfo {author} {\bibfnamefont {T.}~\bibnamefont {Dombeck}}, \bibinfo
  {author} {\bibfnamefont {L.~G.}\ \bibnamefont {Hyman}}, \bibinfo {author}
  {\bibfnamefont {K.}~\bibnamefont {Jaeger}}, \bibinfo {author} {\bibfnamefont
  {B.}~\bibnamefont {Musgrave}}, \bibinfo {author} {\bibfnamefont
  {P.}~\bibnamefont {Schreiner}}, \ and\ \bibinfo {author} {\bibfnamefont
  {R.}~\bibnamefont {Singer~{\it et al.}}},\ }\href@noop {} {\bibfield
  {journal} {\bibinfo  {journal} {Phys.\ Rev.\ D}\ }\textbf {\bibinfo {volume}
  {{\bf 19}}},\ \bibinfo {pages} {2521} (\bibinfo {year} {1979})}\BibitemShut
  {NoStop}%
\bibitem [{\citenamefont {Oset}\ and\ \citenamefont
  {Salcedo}(1987)}]{Oset:1987re}%
  \BibitemOpen
  \bibfield  {author} {\bibinfo {author} {\bibfnamefont {E.}~\bibnamefont
  {Oset}}\ and\ \bibinfo {author} {\bibfnamefont {L.~L.}\ \bibnamefont
  {Salcedo}},\ }\href@noop {} {\bibfield  {journal} {\bibinfo  {journal}
  {Nucl.\ Phys.\ A}\ }\textbf {\bibinfo {volume} {{\bf 468}}},\ \bibinfo
  {pages} {631} (\bibinfo {year} {1987})}\BibitemShut {NoStop}%
\bibitem [{\citenamefont {Bodek}\ and\ \citenamefont
  {Yang}(2002)}]{Bodek:2002vp}%
  \BibitemOpen
  \bibfield  {author} {\bibinfo {author} {\bibfnamefont {A.}~\bibnamefont
  {Bodek}}\ and\ \bibinfo {author} {\bibfnamefont {U.~K.}\ \bibnamefont
  {Yang}},\ }\href@noop {} {\bibfield  {journal} {\bibinfo  {journal} {Nucl.\
  Phys.\ Proc.\ Suppl.}\ }\textbf {\bibinfo {volume} {112}},\ \bibinfo {pages}
  {70} (\bibinfo {year} {2002})}\BibitemShut {NoStop}%
\bibitem [{\citenamefont {Koba}\ \emph {et~al.}(1972)\citenamefont {Koba},
  \citenamefont {Nielsen},\ and\ \citenamefont {Olesen}}]{Koba:1972ng}%
  \BibitemOpen
  \bibfield  {author} {\bibinfo {author} {\bibfnamefont {Z.}~\bibnamefont
  {Koba}}, \bibinfo {author} {\bibfnamefont {H.~B.}\ \bibnamefont {Nielsen}}, \
  and\ \bibinfo {author} {\bibfnamefont {P.}~\bibnamefont {Olesen}},\ }\href
  {\doibase 10.1016/0550-3213(72)90551-2} {\bibfield  {journal} {\bibinfo
  {journal} {Nucl. Phys.}\ }\textbf {\bibinfo {volume} {B40}},\ \bibinfo
  {pages} {317} (\bibinfo {year} {1972})}\BibitemShut {NoStop}%
\bibitem [{\citenamefont {Gran}\ \emph {et~al.}(2013)\citenamefont {Gran},
  \citenamefont {Nieves}, \citenamefont {Sanchez},\ and\ \citenamefont
  {Vicente~Vacas}}]{Gran:2013kda}%
  \BibitemOpen
  \bibfield  {author} {\bibinfo {author} {\bibfnamefont {R.}~\bibnamefont
  {Gran}}, \bibinfo {author} {\bibfnamefont {J.}~\bibnamefont {Nieves}},
  \bibinfo {author} {\bibfnamefont {F.}~\bibnamefont {Sanchez}}, \ and\
  \bibinfo {author} {\bibfnamefont {M.~J.}\ \bibnamefont {Vicente~Vacas}},\
  }\href {\doibase 10.1103/PhysRevD.88.113007} {\bibfield  {journal} {\bibinfo
  {journal} {Phys. Rev.}\ }\textbf {\bibinfo {volume} {D88}},\ \bibinfo {pages}
  {113007} (\bibinfo {year} {2013})},\ \Eprint {http://arxiv.org/abs/1307.8105}
  {arXiv:1307.8105 [hep-ph]} \BibitemShut {NoStop}%
\bibitem [{\citenamefont {Ruiz~Simo}\ \emph {et~al.}(2016)\citenamefont
  {Ruiz~Simo}, \citenamefont {Amaro}, \citenamefont {Barbaro}, \citenamefont
  {De~Pace}, \citenamefont {Caballero}, \citenamefont {Megias},\ and\
  \citenamefont {Donnelly}}]{RuizSimo:2016ikw}%
  \BibitemOpen
  \bibfield  {author} {\bibinfo {author} {\bibfnamefont {I.}~\bibnamefont
  {Ruiz~Simo}}, \bibinfo {author} {\bibfnamefont {J.~E.}\ \bibnamefont
  {Amaro}}, \bibinfo {author} {\bibfnamefont {M.~B.}\ \bibnamefont {Barbaro}},
  \bibinfo {author} {\bibfnamefont {A.}~\bibnamefont {De~Pace}}, \bibinfo
  {author} {\bibfnamefont {J.~A.}\ \bibnamefont {Caballero}}, \bibinfo {author}
  {\bibfnamefont {G.~D.}\ \bibnamefont {Megias}}, \ and\ \bibinfo {author}
  {\bibfnamefont {T.~W.}\ \bibnamefont {Donnelly}},\ }\href@noop {} {\
  (\bibinfo {year} {2016})},\ \Eprint {http://arxiv.org/abs/1607.08451}
  {arXiv:1607.08451 [nucl-th]} \BibitemShut {NoStop}%
\bibitem [{\citenamefont {Sobczyk}(2012)}]{Sobczyk:2012ms}%
  \BibitemOpen
  \bibfield  {author} {\bibinfo {author} {\bibfnamefont {J.~T.}\ \bibnamefont
  {Sobczyk}},\ }\href@noop {} {\bibfield  {journal} {\bibinfo  {journal}
  {Phys.\ Rev.\ C}\ }\textbf {\bibinfo {volume} {86}},\ \bibinfo {pages}
  {015504} (\bibinfo {year} {2012})}\BibitemShut {NoStop}%
\bibitem [{\citenamefont {Schwehr}\ \emph {et~al.}(2016)\citenamefont
  {Schwehr}, \citenamefont {Cherdack},\ and\ \citenamefont
  {Gran}}]{Schwehr:2016pvn}%
  \BibitemOpen
  \bibfield  {author} {\bibinfo {author} {\bibfnamefont {J.}~\bibnamefont
  {Schwehr}}, \bibinfo {author} {\bibfnamefont {D.}~\bibnamefont {Cherdack}}, \
  and\ \bibinfo {author} {\bibfnamefont {R.}~\bibnamefont {Gran}},\ }\href@noop
  {} {\  (\bibinfo {year} {2016})},\ \Eprint {http://arxiv.org/abs/1601.02038}
  {arXiv:1601.02038 [hep-ph]} \BibitemShut {NoStop}%
\bibitem [{\citenamefont {Salcedo}\ \emph {et~al.}(1988)\citenamefont
  {Salcedo}, \citenamefont {Oset}, \citenamefont {Vicente-Vacas},\ and\
  \citenamefont {Garcia-Recio}}]{Salcedo:1987md}%
  \BibitemOpen
  \bibfield  {author} {\bibinfo {author} {\bibfnamefont {L.~L.}\ \bibnamefont
  {Salcedo}}, \bibinfo {author} {\bibfnamefont {E.}~\bibnamefont {Oset}},
  \bibinfo {author} {\bibfnamefont {M.~J.}\ \bibnamefont {Vicente-Vacas}}, \
  and\ \bibinfo {author} {\bibfnamefont {C.}~\bibnamefont {Garcia-Recio}},\
  }\href@noop {} {\bibfield  {journal} {\bibinfo  {journal} {Nucl.\ Phys.\ A}\
  }\textbf {\bibinfo {volume} {484}},\ \bibinfo {pages} {557} (\bibinfo {year}
  {1988})}\BibitemShut {NoStop}%
\bibitem [{\citenamefont {Benhar}\ and\ \citenamefont
  {Meloni}(2007)}]{Benhar:2006nr}%
  \BibitemOpen
  \bibfield  {author} {\bibinfo {author} {\bibfnamefont {O.}~\bibnamefont
  {Benhar}}\ and\ \bibinfo {author} {\bibfnamefont {D.}~\bibnamefont
  {Meloni}},\ }\href {\doibase 10.1016/j.nuclphysa.2007.02.015} {\bibfield
  {journal} {\bibinfo  {journal} {Nucl. Phys.}\ }\textbf {\bibinfo {volume}
  {A789}},\ \bibinfo {pages} {379} (\bibinfo {year} {2007})},\ \Eprint
  {http://arxiv.org/abs/hep-ph/0610403} {arXiv:hep-ph/0610403 [hep-ph]}
  \BibitemShut {NoStop}%
\end{thebibliography}%

\end{document}